\documentclass[twocolumn, superscriptaddress, prb, showpacs]{revtex4-1}

\usepackage{graphicx}

\newcommand{\NH}{NH$_3$BH$_3$}
\newcommand{\arialbf}{\bfseries\sffamily}

\begin{document}

\title{Positional disorder in ammonia\\
borane at ambient conditions}

\author{E. Welchman}
\affiliation{Department of Physics, Wake Forest University,
Winston-Salem, NC 27109, USA}

\author{P. Giannozzi}
\affiliation{Department of Chemistry, Physics, and Environment,
University of Udine, 33100 Udine, Italy}

\author{T. Thonhauser}\email{thonhauser@wfu.edu}
\affiliation{Department of Physics, Wake Forest University,
Winston-Salem, NC 27109, USA}

\date{\today}

\begin{abstract}
We solve a long-standing experimental discrepancy of \NH, which---as a
molecule---has a three-fold rotational axis, but in its crystallized
form at room temperature shows a four-fold symmetry around the same
axis, creating a geometric incompatibility.  To explain this peculiar
experimental result, we study the dynamics of this system with \emph{ab
initio} Car-Parrinello molecular dynamics and nudged-elastic band
simulations. We find that rotations, rather than spatial static
disorder, at angular velocities of 2 rev/ps---a time-scale too small to
be resolved by standard experimental techniques---are responsible for
the four-fold symmetry.
\end{abstract}

\maketitle

\begin{figure}
\centering\includegraphics[width={.7\columnwidth}]{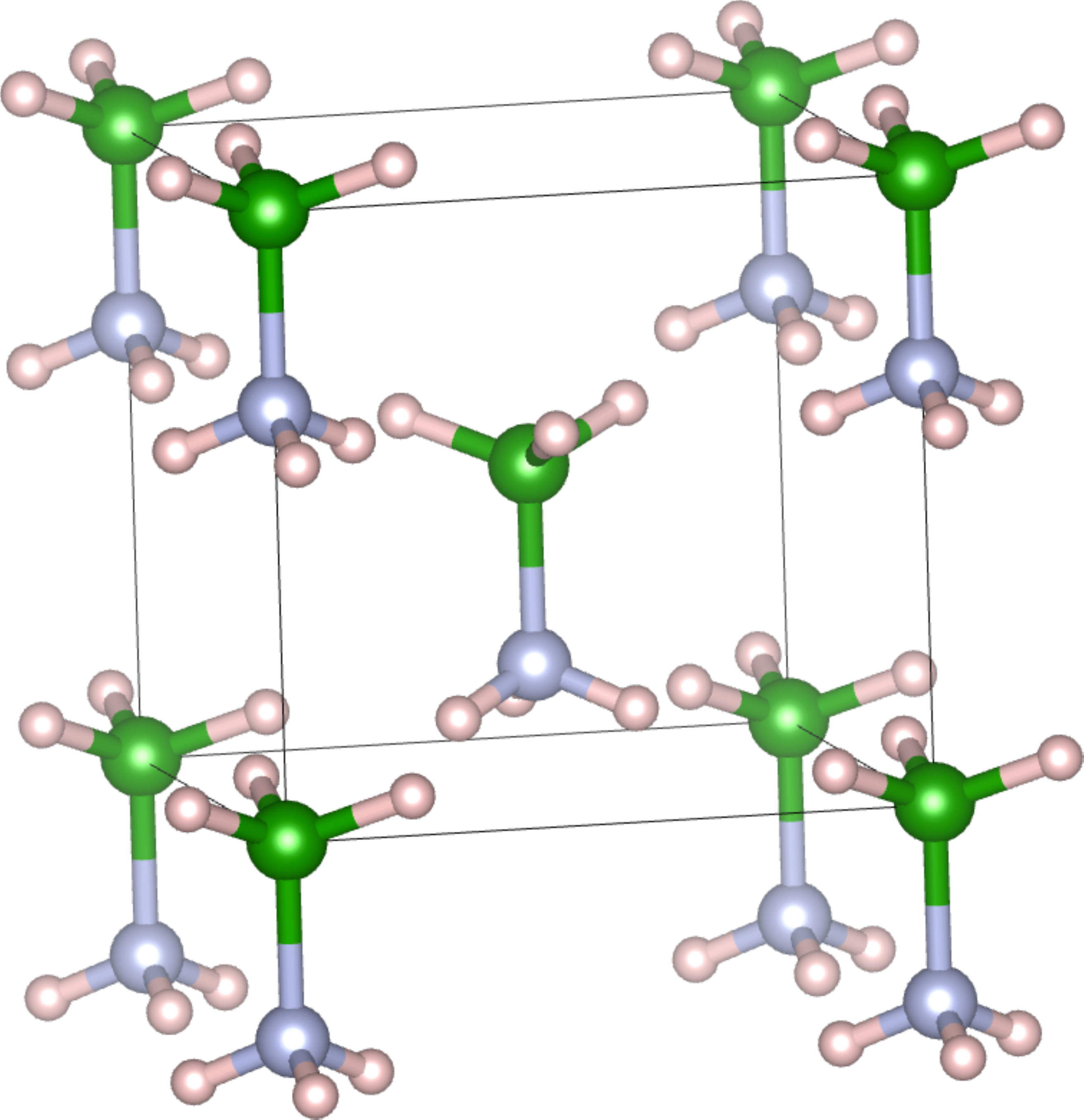}
\caption{\label{fig:structure}Structure of the high-temperature,
body-centered tetragonal phase of NH$_3$BH$_3$. Note that the locations
of H atoms in this figure are not indicative of experimental results,
but just one possibility of how the halos could be oriented in the
solid.}
\end{figure}

Ammonia borane NH$_{3}$BH$_{3}$ has drawn significant interest in recent
years because of its potential as a hydrogen storage material, with a
gravimetric storage density of 19.6
mass\%.\cite{Xiong_2008:high-capacity_hydrogen,
Chua_2011:development_amidoboranes, Swinnen_2010:potential_hydrogen,
Hamilton_2009:b-n_compounds, Heldebrant_2008:effects_chemical,
Marder_2007:will_we, Kim_2009:determination_structure} The structure of
its solid phase has been explored
previously,\cite{Reynhardt_1983:molecular_dynamics,
Penner_1999:deuterium_nmr, Klooster_1999:study_n-hh-b,
Brown_2006:dynamics_ammonia, Bowden_2007:room-temperature_structure,
Lin_2012:experimental_theoretical} but the literature does not agree
about the hydrogen behavior at room temperature. The molecule consists
of a dative B--N bond and a trio of H atoms (henceforth referred to as a
`halo') bonded to each of those two atoms, forming an hourglass shape,
visible in Fig.~\ref{fig:structure}. At low temperatures (0 $\sim$
225~K), the solid exhibits an orthorhombic structure with space group
\emph{Pmn}2$_{1}$. Heated above 225~K, it undergoes a phase transition
to a body-centered tetragonal structure with space group
\emph{I}4\emph{mm}. It is this room-temperature phase that exhibits
unexpected experimental results: while the molecule itself has a
three-fold symmetry about the B--N axis,
neutron\cite{Brown_2006:dynamics_ammonia, Kumar_2010:pressure_induced}
and X-ray\cite{Filinchuk_2009:high-pressure_phase, Chen_2010:situ_x-ray,
Bowden_2007:room-temperature_structure} diffraction on the solid reveal
a four-fold symmetry about the same axis, creating a geometric
incompatibility within the structure.  Investigating the dynamics of the
system with \emph{ab initio} methods, we find that the individual halos
are rotating with angular velocity on the order of 0.7~deg/fs $\approx$
2 rev/ps, such that standard experiments can only probe the time
averaged positions, leading to the tetragonal host structure with
four-fold symmetry.

The precise behavior of these hydrogen halos has been the subject of
several studies over three decades. In 1983, Reynhardt and
Hoon\cite{Reynhardt_1983:molecular_dynamics} found three-fold
reorientations of the BH$_{3}$ and NH$_{3}$ groups with a tunneling
frequency of 1.4~MHz in the orthorhombic phase. Penner et
al.\cite{Penner_1999:deuterium_nmr} found in 1999 that these groups
reoriented independently. Deciphering the behavior in the tetragonal
structure has been less straightforward. In the same 1983 study,
Reynhardt and Hoon concluded that the BH$_{3}$, and possibly the
NH$_{3}$ groups, rotate freely. Brown et
al.\cite{Brown_2006:dynamics_ammonia} found that they could describe the
disorder entirely with three-fold jump diffusion. Bowden et al.\ tried
using a larger unit cell to model the same disorder as spatial variation
rather than higher-order rotation; however, they found no evidence to
support this model,\cite{Bowden_2007:room-temperature_structure} leaving
this disagreement unresolved in the literature.  The present study aims
to elucidate how the hydrogen halos behave in the solid, especially in
the high-temperature, tetragonal structure. To this end, we find thermal
barriers to rotation in gas phase as well as both orthorhombic and
tetragonal phases. We supplement these findings with \emph{ab initio}
molecular dynamics simulations to track individual halos' behavior.

Our \emph{ab initio} simulations are at the density functional theory
level, using a plane-wave basis.  Since ammonia borane is a strong van
der Waals complex,\cite{Chen_2010:situ_x-ray,
Lin_2008:raman_spectroscopy} the inclusion of van der Waals forces is
essential;\cite{Klooster_1999:study_n-hh-b,
Wolstenholme_2011:homopolar_dihydrogen,
Wolstenholme_2012:thermal_desorption} we thus use
vdW-DF1\cite{Dion_2004:van_waals, Thonhauser_2007:van_waals,
Langreth_2009:density_functional} (i.e.\ revPBE exchange and LDA
correlation in addition to the nonlocal contribution) as the
exchange-correlation functional for all calculations.  Car-Parrinello
molecular dynamics (CPMD) was performed with the CP code (part of
\textsc{Quantum-Espresso} version 5.0.2; the vdW-DF capability in CP is
a new feature, which we have just
implemented),\cite{Giannozzi_2009:quantum_espresso} using ultrasoft
pseudopotentials and wave function and density cutoffs of 475 and 5700
eV. The CPMD simulations used an electronic convergence of 10$\time
10^{-8}$~eV, a fictitious electron mass of 400~a.u., and a time step of
5 a.u. We further used a $2\times2\times2$ supercell, accommodating
16 molecules, and started from the experimental lattice
constants\cite{Bowden_2007:room-temperature_structure} at
297 and 90~K for tetragonal and orthorhombic phases, respectively.
Similar calculations have been done
previously,\cite{Liang_2012:first-principles_study} but with a different
functional and at much higher temperature.  Climbing image
nudged-elastic band (NEB) simulations to find precise rotational
barriers for halos and entire molecules were performed with
\textsc{Vasp} (version 5.3.3),\cite{Kresse_1996:efficient_iterative,
Kresse_1999:ultrasoft_pseudopotentials} utilizing PAW potentials and a
cutoff of 500~eV. For solid phase barrier calculations, we used a
$4\times4\times4$ $k$-point mesh and 8 images for NEB calculations. In
the gas phase, we used only the gamma point and 16 images.  Note that
nuclear quantum effects have not been taken into account. Further
information, in particular including structural information for all our
simulations, can be found in the Supplementary Materials.


We begin by investigating the barriers for rotations in different
situations, i.e.\ the gas-phase molecule and the orthorhombic and
tetragonal solid phases.  Depending on the situation, we performed two
kinds of simulations: ``fixed'' labels simulations where the geometry of
the cell as well as the halo has been fixed and the entire halo or
molecule is rotated around the axis in a rigid manner. ``NEB'' refers to
the transition state formalism of the nudged-elastic band method, where
the geometry of the halo can change and adapt along the path, allowing
it to lower its energy. While the latter is preferable due to its higher
accuracy for barriers, we also use the former i) in order to compare to
previous quantum-chemistry calculations; and ii) for the tetragonal
high-temperature phase, in which NEB leads to unphysical deformations,
as all DFT ground-state simulations are technically done at 0~K and the
structure attempts to mimic the orthorhombic phase. Results are
summarized in Table~\ref{tab:Torsion_results} and detailed curves for
the barriers can be found in the Supplementary Materials.

The situation in the gas-phase molecule is the simplest. Completing a
fixed rotation of one halo results in a thermal barrier of 84.7~meV,
within 5\% of an empirical estimate\cite{Thorne_1983:microwave_spectrum}
and in very good agreement with quantum-chemistry
calculations,\cite{Demaison_2008:equilibrium_structure} validating our
methodology. NEB calculations necessarily decrease the estimate of the
barrier, in this case yielding a value of 79.1~meV.

In a crystalline environment, dihydrogen bonds between molecules affect
how each molecule behaves. In the orthorhombic phase, the dihydrogen
bond network creates a 67.5~meV barrier to rotating the entire molecule
(NEB). This barrier is low enough that molecules in a crystal can
reorient at some rate, given enough temperature. It is interesting to
see that a calculation with fixed halo geometry results in a much higher
barrier, attesting to the fact that the rotating halo and its
surroundings prefer to undergo significant deformation and reorientation
during the rotation.  For instance, the orientation of the B-N axis
prefers to precess as the B halo is rotated in an attempt to maximize
the strength of dihydrogen bonds with its neighbors. 

The ease of the rotation process is dependent on which individual halo
is rotated.  Our calculations for the N halo barriers are in good
agreement with experimental findings (summarized in
Table~\ref{tab:Torsion_results}).  Accuracy for the B halo is more
difficult to gauge. Our results for a fixed rotation are in agreement
with a previous theoretical
study,\cite{Parvanov_2008:materials_hydrogen} but experimental values
line up almost exactly halfway between our calculations for fixed
rotation and NEB barriers. Regardless of the magnitude of the
difference, we find that the BH$_3$ group faces a larger barrier to
rotation than the NH$_3$ group, in agreement with the literature.

\begin{table}\small
\caption{\label{tab:Torsion_results}Numerical values for calculated
rotational barriers in meV and values given in the literature. Error
bars for experimental values in the literature typically range from
5 to 10~meV.}
\begin{tabular*}{\columnwidth}{@{}@{\extracolsep{\fill}}lrrr@{}}              \hline\hline
                       & fixed & NEB   & literature                            \\\hline
\multicolumn{4}{@{}l}{\arialbf gas phase}                                      \\
one halo               &  84.7 &  79.1 & 89.8,$^{a*}$ 86.7$^{b\ddag}$                    \\[1.5ex]
\multicolumn{4}{@{}l}{\arialbf orthorhombic phase}                             \\
N halo                 & 106.6 &  94.9 & 100,$^{c*}$ 142,$^{d*}$ 82.7,$^{e*}$ 131.6$^{f\dag\ddag}$ \\
B halo                 & 443.9 & 102.9 & 260,$^{c*}$ 259,$^{d*}$ 397$^{f\dag}$             \\
molecule               & 403.4 &  67.5 &     328$^{f\dag}$                           \\[1.5ex]
\multicolumn{4}{@{}l}{\arialbf tetragonal phase}                               \\
N halo                 &  60.1 &       & 75.7,$^{d*}$ 50.8$^{e*}$                    \\
B halo                 &  61.9 &       & 60.8,$^{c*}$ 61.1,$^{d*}$ 50.8$^{e*}$          \\
molecule               &  19.4 &       &                                       \\\hline\hline
\end{tabular*}
$^a$Ref. \citenum{Thorne_1983:microwave_spectrum},
$^b$Ref. \citenum{Demaison_2008:equilibrium_structure},
$^c$Ref. \citenum{Reynhardt_1983:molecular_dynamics},
$^d$Ref. \citenum{Penner_1999:deuterium_nmr},
$^e$Ref. \citenum{Brown_2006:dynamics_ammonia},
$^f$Ref. \citenum{Parvanov_2008:materials_hydrogen}\\
$^\dag$DFT (B3LYP),
$^\ddag$quantum chemistry,
$^*$experiment
\end{table}

In the high-temperature tetragonal phase, the rotation of either halo
has essentially the same barrier of $\sim$61~meV, in good agreement with
the literature (again see Table~\ref{tab:Torsion_results}). But, even
more important, rotating the entire molecule requires just 19.4~meV.
This barrier is easily overcome at ambient conditions ($k_BT= 25$~meV at
room temperature). Previous studies have argued that NH$_{3}$ and
BH$_{3}$ groups rotate freely\cite{Reynhardt_1983:molecular_dynamics}
and that the molecule rotates as a
whole.\cite{Penner_1999:deuterium_nmr} Our evidence supports a
combination of both explanations. The barrier to rotating the whole
molecule is low enough that it can occur freely at room temperature. The
torsional barrier for each group also allows them to rotate
independently; since these barriers are within 2~meV, this rate should
be equivalent between the groups, leading atoms in both groups to move
at the same rate, as seen
experimentally.\cite{Brown_2006:dynamics_ammonia} Based on the barriers
in Table~\ref{tab:Torsion_results}, from the Arrhenius equation we
estimate (assuming the same pre-exponential factor for all rotations)
that whole-molecule rotation occurs at about five times the rate of
individual halo rotations at room temperature and approximately
20 and 30 times the rate for rotating individual halos in the
low-temperature phase.

Also of note is that torsional barriers in the orthorhombic phase are
larger than those of an isolated molecule, whereas the torsional
barriers in the tetragonal phase are lower. This result alone shows that
in the low-temperature phase rotation is suppressed, while it is
encouraged in the high-temperature phase.

\begin{figure}
\centering\includegraphics[width=\columnwidth]{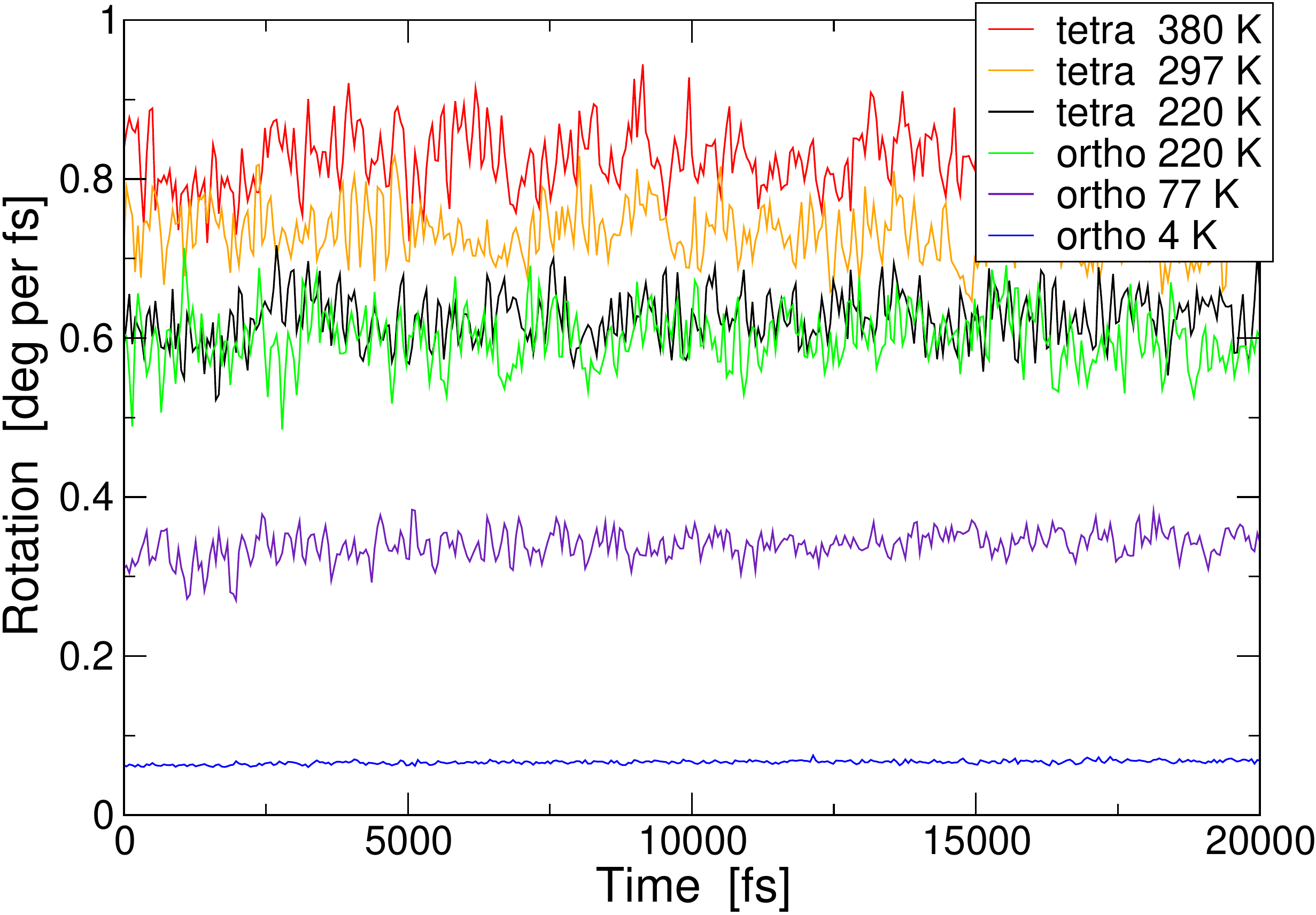}
\caption{\label{fig:avgomega}Average angular velocity among H atoms at
each frame in the simulation. The plot shows a running average over 50
frames. Note that this analysis only captures the angular
velocity---motion of the single hydrogen atoms in a direction parallel
to the B--N bond is not captured here, making up for some of the
``missing'' kinetic energy and keeping the temperature constant.}
\end{figure}

With the knowledge of the barriers, we now move to the analysis of the
dynamics of the crystalline phase. We performed CPMD simulations in both
the orthorhombic (at 4, 77, and 220 K) and tetragonal (at 220, 297, and
380~K) phases in order to study the motion of H atoms in the NH$_{3}$
and BH$_{3}$ groups. We used 1~ps for thermalization of the system and
thereafter performed 20~ps production runs. Analyzing the corresponding
trajectories leads to the initial (obvious) conclusion that halos rotate
more rapidly at higher temperatures.

To substantiate this claim, we calculated for each H atom in the
simulation the angular velocity about the nearest B--N axis. We then
averaged the absolute value of this angular velocity---otherwise there
is a lot of cancellation, as halos rotate in both directions---over all
H atoms in the simulation to measure how rapidly the halos are rotating
in each frame of the simulation. The results of this calculation are
shown in Fig.~\ref{fig:avgomega}, confirming the idea that H atoms
rotate more quickly at higher temperatures and giving quantitative
values for the speed of their rotation, in qualitative agreement with
the barriers found earlier. It is important to note that in simulating
the tetragonal supercell, the B--N axes typically maintain an
instantaneous tilt between 5 and 20 degrees from vertical. At 297 and
380~K, the average orientation is vertical, whereas at 220~K there
appears to be a correlation between neighbors similar to that found in
the low-temperature phase. Consequently, the dynamics of the 220~K
simulations are qualitatively very similar.

\begin{figure}
\centering
\includegraphics[width=0.8\columnwidth]{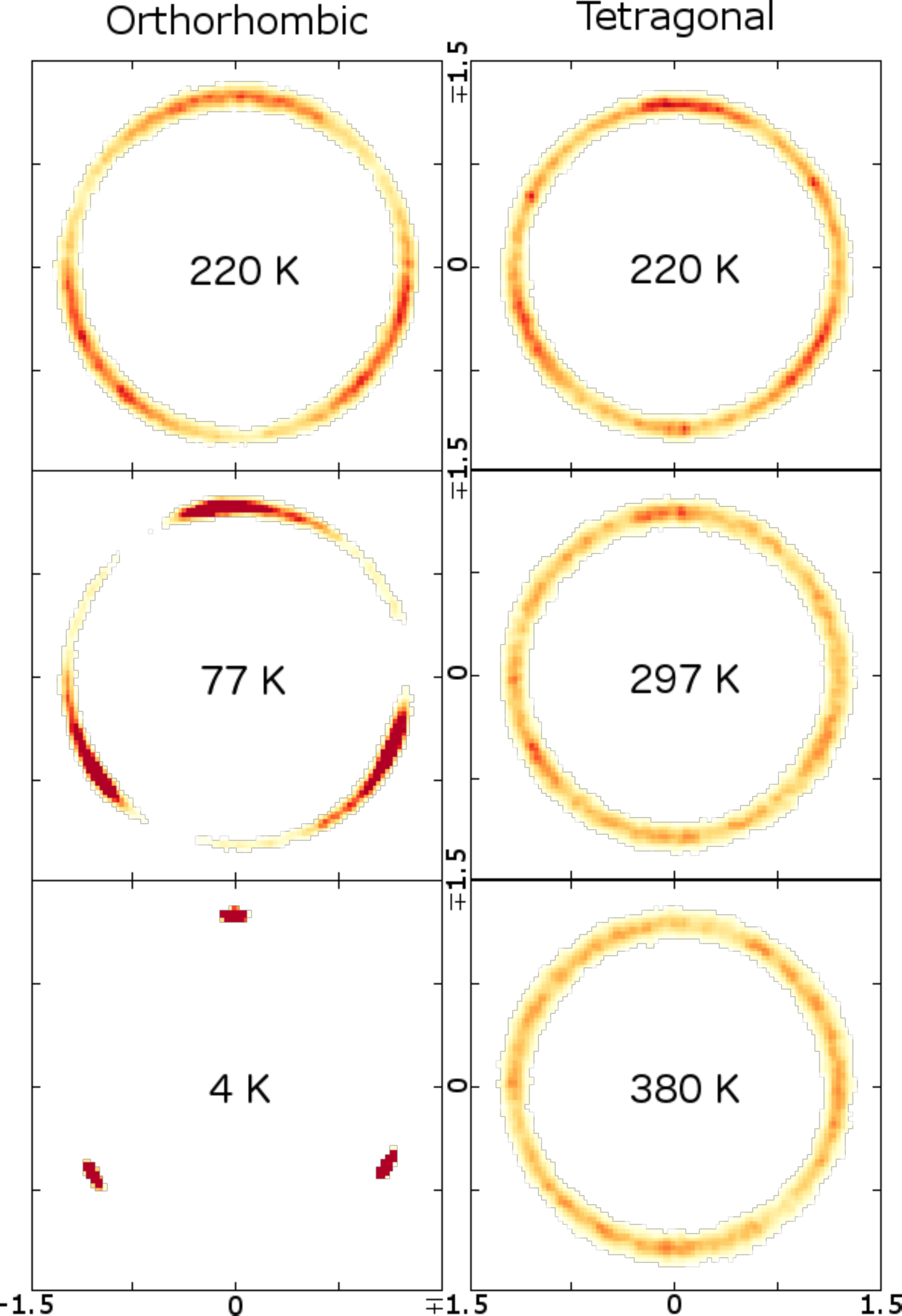}
\caption{\label{fig:heatmaps}Heat maps of the location of all three H
atoms in one N-group halo over the course of CPMD simulations. The
corresponding B-group heat maps look very similar. Positions have been
flattened into a plane perpendicular to the B--N axis. Motion along this
axis is not apparent in these plots. Each map is 3~\AA$^2$.}
\end{figure}

The order of magnitude for rotations is found to be 0.7~deg/fs $\approx$
2 rev/ps at room temperature. As such, halos can easily rotate 120 deg
in 0.2 ps. Unless experiments can be performed with a resolution smaller
than this, they will see time-averaged positions, and halos (with a
three-fold symmetry) start looking like rings, as described below.
Experiments will thus pick up the symmetry of the tetragonal host
lattice, explaining the four-fold symmetry.

Casual observation of the simulations reveals that halos in the
high-temperature phase are unlikely to undergo full revolutions in a
short burst. Rather, a more accurate description of the qualitative
behavior is that---as a halo moves close to a neighboring halo---they
will rotate some amount in order to form a dihydrogen bond. The halo
will then move closer to a different neighbor and adopt a different
alignment. A halo equally far from all of its neighbors will also follow
the realignment of the opposite halo of the same molecule. Because the
molecules are constantly oscillating in the crystal structure due to
thermal energy, these reorientation processes result in a constantly
shifting dihydrogen bond network.

This analysis above describes how rapidly H atoms rotate about their
native molecules, but does not describe where they are.  To give a more
systematic estimate of hydrogen position over time, we provide ``heat
maps'' in Fig.~\ref{fig:heatmaps} that describe what angular positions
the H atoms inhabit over the course of a whole simulation. Each heat map
shows the occupation density for all three H atoms in a particular halo.
These heat maps demonstrate a clear pattern of increasing positional
disorder at higher temperature. The three-fold symmetry inherent in the
molecular structure is apparent in the maps from the orthorhombic phase.
This symmetry becomes much less clear in the tetragonal phase,
indicating that rotation represents a significant source of the disorder
found experimentally. Furthermore, the occupational density in the
higher-temperature structure is much more spread out angularly,
indicating that reorientation is not limited to 120-degree jumps, as
concluded by Brown et al.,\cite{Brown_2006:dynamics_ammonia} but is a
more fluid process.


In summary, we have calculated torsional and rotational barriers for
NH$_{3}$BH$_{3}$ in the gas phase and both low- and high-temperature
crystalline structures.  In addition, we have studied the dynamics of
the crystalline phase explicitly with CPMD simulations. Our calculations
indicate that in the low-temperature orthorhombic phase, the BH$_{3}$
and NH$_{3}$ groups reorient along a three-fold rotational potential at
different rates. Both entire-molecule and independent reorientations
contribute to the experimental rates found previously.  In the
high-temperature tetragonal phase, on the other hand, the barrier to
entire-molecule rotation is low enough that thermal energy in ambient
conditions allows the molecule to overcome the three-fold rotational
potential. Consequently, the molecule is able to rotate freely with
angular velocities on the order of 2 rev/ps. By quantifying the speed of
those rotations, we thus resolve a long-standing experimental
discrepancy, where a molecule with three-fold symmetry shows four-fold
symmetry around the same axis in its crystalline form.

This work was supported in full by NSF Grant No.\ DMR-1145968.

\bibliography{references}

\end{document}


\title{Positional disorder in ammonia\\
borane at ambient conditions\\[1ex]
--- Supplemental Material ---\\}

\author{E. Welchman}
\affiliation{Department of Physics, Wake Forest University,
Winston-Salem, NC 27109, USA}

\author{P. Giannozzi}
\affiliation{Department of Chemistry, Physics, and Environment,
University of Udine, 33100 Udine, Italy}

\author{T. Thonhauser}\email{thonhauser@wfu.edu}
\affiliation{Department of Physics, Wake Forest University,
Winston-Salem, NC 27109, USA}

\date{\today}
\maketitle

\section{Optimized Structures}

\subsection{Isolated Molecule}

\noindent Cartesian coordinates (\emph{x}, \emph{y}, \emph{z}) in
\text{\AA}, for the optimized structure found by our simulations of an
isolated NH$_3$BH$_3$ molecule.
\begin{center}\scriptsize
\begin{verbatim}
     B      0.0000000   0.0000000   0.0000000
     N      0.0000000   1.6856212   0.0000000
     H      0.9519811   2.0563251   0.0000000
     H     -1.1740483  -0.3022489   0.0000000
     H     -0.4759065   2.0567762   0.8242625
     H     -0.4759065   2.0567762  -0.8242625
     H      0.5870187  -0.3021106  -1.0167601
     H      0.5870187  -0.3021106   1.0167602
\end{verbatim}
\end{center}

\subsection{Orthorhombic Structure}

\noindent Lattice parameters and fractional coordinates ($x/a$, $y/b$,
and $z/c$) resulting from our simulations optimizing the structure of
NH$_3$BH$_3$ in the low-temperature, orthorhombic phase.
\begin{eqnarray*}
a      &=& 5.44233\;\text{\AA}\\
b      &=& 4.94048\;\text{\AA}\\
c      &=& 5.13446\;\text{\AA}\\
\alpha &=& \beta = \gamma = 90.0^{\circ}\\
\end{eqnarray*}
\begin{center}\scriptsize
\begin{verbatim}
     B      0.0000000   0.1830125   0.0019077
     B      0.5000000   0.8169874   0.5019076
     N      0.0000000   0.2402808   0.3127204
     N      0.5000000   0.7597191   0.8127204
     H      0.0000000   0.4427503   0.3555205
     H      0.5000000   0.5572496   0.8555205
     H      0.1520232   0.1583956   0.4010154
     H      0.3479767   0.8416043   0.9010154
     H      0.6520232   0.8416043   0.9010154
     H      0.8479767   0.1583956   0.4010154
     H      0.0000000   0.9368787   0.9737231
     H      0.5000000   0.0631212   0.4737231
     H      0.1851884   0.2836140   0.9105986
     H      0.3148115   0.7163859   0.4105986
     H      0.6851884   0.7163859   0.4105986
     H      0.8148115   0.2836140   0.9105986
\end{verbatim}
\end{center}

\newpage\subsection{Tetragonal Structure --- From Experiment}

\noindent Lattice parameters and fractional coordinates ($x/a$, $y/b$,
and $z/c$) of the high-temperature, tetragonal structure of NH$_3$BH$_3$,
taken from experiment and refined in space group $I4mm$. Taken directly
from Ref.~\citenum{Bowden_2007:room-temperature_structure}.
\begin{eqnarray*}
a      &=& b = 5.2630\;\text{\AA}\\
c      &=&  5.0504\;\text{\AA}\\
\end{eqnarray*}
\begin{center}\scriptsize
\begin{verbatim}
     B      0.0000   0.0000   0.0032
     N      0.0000   0.0000   0.6869
     H      0.0000   0.1480   0.6190
     H      0.0000   0.1990   0.0770
\end{verbatim}
\end{center}

\subsection{Tetragonal Structure --- For Calculations}

\noindent Lattice parameters and fractional coordinates ($x/a$, $y/b$,
and $z/c$) used for the high-temperature, tetragonal structure of
NH$_3$BH$_3$. The structure from
Ref.~\citenum{Bowden_2007:room-temperature_structure} was used as a
starting point. Hydrogen halos were completed to include twelve atoms in
order to demonstrate both four-fold and three-fold rotational
symmetries. We then found the energy of each possible arrangement of the
three-atom halos out of these available positions where each molecule
retained its characteristic three-fold symmetry. The arrangement with
the lowest energy is shown here:
\begin{eqnarray*}
a      &=& b = 5.2630\;\text{\AA}\\
c      &=&  5.0504\;\text{\AA}\\
\end{eqnarray*}
\begin{center}\scriptsize
\begin{verbatim}
     B      0.0000000  0.0000000  0.0032000
     B      0.5000000  0.5000000  0.5032000
     N      0.0000000  0.0000000  0.6869000
     N      0.5000000  0.5000000  0.1869000
     H      0.8010000  0.0000000  0.0770000
     H      0.0995000  0.1723390  0.0770000
     H      0.0995000  0.8276610  0.0770000
     H      0.1480000  0.0000000  0.6190000
     H      0.9260000  0.8718280  0.6190000
     H      0.9260000  0.1281720  0.6190000
     H      0.4005000  0.6723390  0.5770000
     H      0.6990000  0.5000000  0.5770000
     H      0.4005000  0.3276610  0.5770000
     H      0.5740000  0.3718282  0.1190000
     H      0.3519998  0.5000000  0.1190000
     H      0.5740002  0.6281719  0.1190000
\end{verbatim}
\end{center}

\newpage\section{Molecular Dynamics Simulations}

\noindent For our molecular dynamics calculations, we used the CP code
(part of \textsc{Quantum-Espresso}), utilizing ultra-soft
pseudopotentials and wave function and density cutoffs of 475 and 5700
eV with an electronic convergence of 10$\time 10^{-8}$~eV, a fictitious
electron mass of 400~a.u., and a time step of 5 a.u. We used a Nos\'{e}
thermostat with an oscillation frequency of 6.5~THz and allowed the
simulations to thermalize for at least 1~ps. Production runs were
typically at least 20 ps. The starting structures are listed below.

\subsection{Orthorhombic Structure}

\noindent Starting fractional coordinates ($x/a$, $y/b$, and $z/c$) and
lattice parameters used for CPMD calculations in the low-temperature,
orthorhombic phase. A $2\times 2\times 2$ supercell was used, including
16 NH$_3$BH$_3$ molecules.
\begin{eqnarray*}
a      &=& 10.88466\;\text{\AA}\\
b      &=& 9.88096\;\text{\AA}\\
c      &=& 10.26892\;\text{\AA}\\
\alpha &=& \beta = \gamma = 90.0^{\circ}\\
\end{eqnarray*}
\vspace{-6ex}
\begin{center}\scriptsize
\begin{verbatim}
     B      0.0000000  0.0804724  0.0009784
     B      0.0000000  0.0804724  0.4901925
     B      0.0000000  0.5566408  0.0009784
     B      0.0000000  0.5566408  0.4901925
     B      0.5090651  0.0804724  0.0009784
     B      0.5090651  0.0804724  0.4901925
     B      0.5090651  0.5566408  0.0009784
     B      0.5090651  0.5566408  0.4901925
     B      0.2545325  0.3956959  0.2455854
     B      0.2545325  0.3956959  0.7347996
     B      0.2545325  0.8718642  0.2455854
     B      0.2545325  0.8718642  0.7347996
     B      0.7635976  0.3956959  0.2455854
     B      0.7635976  0.3956959  0.7347996
     B      0.7635976  0.8718642  0.2455854
     B      0.7635976  0.8718642  0.7347996
     N      0.0000000  0.1170898  0.1526837
     N      0.0000000  0.1170898  0.6418978
     N      0.0000000  0.5932580  0.1526837
     N      0.0000000  0.5932580  0.6418978
     N      0.5090651  0.1170898  0.1526837
     N      0.5090651  0.1170898  0.6418978
     N      0.5090651  0.5932580  0.1526837
     N      0.5090651  0.5932580  0.6418978
     N      0.2545325  0.3590785  0.3972908
     N      0.2545325  0.3590785  0.8865048
     N      0.2545325  0.8352468  0.3972908
     N      0.2545325  0.8352468  0.8865048
     N      0.7635976  0.3590785  0.3972908
     N      0.7635976  0.3590785  0.8865048
     N      0.7635976  0.8352468  0.3972908
     N      0.7635976  0.8352468  0.8865048
     H      0.0000000  0.1971337  0.1629083
     H      0.0000000  0.1971337  0.6521223
     H      0.0000000  0.6733020  0.1629083
     H      0.0000000  0.6733020  0.6521223
     H      0.5090651  0.1971337  0.1629083
     H      0.5090651  0.1971337  0.6521223
     H      0.5090651  0.6733020  0.1629083
     H      0.5090651  0.6733020  0.6521223
     H      0.2545325  0.2790346  0.4075153
     H      0.2545325  0.2790346  0.8967295
     H      0.2545325  0.7552029  0.4075153
     H      0.2545325  0.7552029  0.8967295
     H      0.7635976  0.2790346  0.4075153
     H      0.7635976  0.2790346  0.8967295
     H      0.7635976  0.7552029  0.4075153
     H      0.7635976  0.7552029  0.8967295
     H      0.0717782  0.0795201  0.1932396
     H      0.0717782  0.0795201  0.6824537
     H      0.0717782  0.5556884  0.1932396
     H      0.0717782  0.5556884  0.6824537
     H      0.5808432  0.0795201  0.1932396
     H      0.5808432  0.0795201  0.6824537
     H      0.5808432  0.5556884  0.1932396
     H      0.5808432  0.5556884  0.6824537
     H      0.1827544  0.3966482  0.4378466
     H      0.1827544  0.3966482  0.9270607
     H      0.1827544  0.8728165  0.4378466
     H      0.1827544  0.8728165  0.9270607
     H      0.6918194  0.3966482  0.4378466
     H      0.6918194  0.3966482  0.9270607
     H      0.6918194  0.8728165  0.4378466
     H      0.6918194  0.8728165  0.9270607
     H      0.3263107  0.3966482  0.4378466
     H      0.3263107  0.3966482  0.9270607
     H      0.3263107  0.8728165  0.4378466
     H      0.3263107  0.8728165  0.9270607
     H      0.8353757  0.3966482  0.4378466
     H      0.8353757  0.3966482  0.9270607
     H      0.8353757  0.8728165  0.4378466
     H      0.8353757  0.8728165  0.9270607
     H      0.4372869  0.0795201  0.1932396
     H      0.4372869  0.0795201  0.6824537
     H      0.4372869  0.5556884  0.1932396
     H      0.4372869  0.5556884  0.6824537
     H      0.9463518  0.0795201  0.1932396
     H      0.9463518  0.0795201  0.6824537
     H      0.9463518  0.5556884  0.1932396
     H      0.9463518  0.5556884  0.6824537
     H      0.0000000  0.4428365  0.4862788
     H      0.0000000  0.4428365  0.9754928
     H      0.0000000  0.9190048  0.4862788
     H      0.0000000  0.9190048  0.9754928
     H      0.5090651  0.4428365  0.4862788
     H      0.5090651  0.4428365  0.9754928
     H      0.5090651  0.9190048  0.4862788
     H      0.5090651  0.9190048  0.9754928
     H      0.2545325  0.0333318  0.2416718
     H      0.2545325  0.0333318  0.7308858
     H      0.2545325  0.5095001  0.2416718
     H      0.2545325  0.5095001  0.7308858
     H      0.7635976  0.0333318  0.2416718
     H      0.7635976  0.0333318  0.7308858
     H      0.7635976  0.5095001  0.2416718
     H      0.7635976  0.5095001  0.7308858
     H      0.0789051  0.1309463  0.4490985
     H      0.0789051  0.1309463  0.9383126
     H      0.0789051  0.6071146  0.4490985
     H      0.0789051  0.6071146  0.9383126
     H      0.5879701  0.1309463  0.4490985
     H      0.5879701  0.1309463  0.9383126
     H      0.5879701  0.6071146  0.4490985
     H      0.5879701  0.6071146  0.9383126
     H      0.1756274  0.3452220  0.2044915
     H      0.1756274  0.3452220  0.6937056
     H      0.1756274  0.8213903  0.2044915
     H      0.1756274  0.8213903  0.6937056
     H      0.6846925  0.3452220  0.2044915
     H      0.6846925  0.3452220  0.6937056
     H      0.6846925  0.8213903  0.2044915
     H      0.6846925  0.8213903  0.6937056
     H      0.3334376  0.3452220  0.2044915
     H      0.3334376  0.3452220  0.6937056
     H      0.3334376  0.8213903  0.2044915
     H      0.3334376  0.8213903  0.6937056
     H      0.8425027  0.3452220  0.2044915
     H      0.8425027  0.3452220  0.6937056
     H      0.8425027  0.8213903  0.2044915
     H      0.8425027  0.8213903  0.6937056
     H      0.4301600  0.1309463  0.4490985
     H      0.4301600  0.1309463  0.9383126
     H      0.4301600  0.6071146  0.4490985
     H      0.4301600  0.6071146  0.9383126
     H      0.9392250  0.1309463  0.4490985
     H      0.9392250  0.1309463  0.9383126
     H      0.9392250  0.6071146  0.4490985
     H      0.9392250  0.6071146  0.9383126
\end{verbatim}
\end{center}

\newpage\subsection{Tetragonal Structure}

\noindent Starting fractional coordinates ($x/a$, $y/b$, and $z/c$) and
lattice parameters used for CPMD calculations in the high-temperature,
tetragonal phase.
\begin{eqnarray*}
a   &=& b = 10.526\;\text{\AA}\\
c   &=&  10.1008\;\text{\AA}\\
\end{eqnarray*}
\begin{center}\scriptsize
\begin{verbatim}
     B      0.0000000  0.0000000  0.0016201
     B      0.0000000  0.0000000  0.5016280
     B      0.0000000  0.5000038  0.0016201
     B      0.0000000  0.5000038  0.5016280
     B      0.5000038  0.0000000  0.0016201
     B      0.5000038  0.0000000  0.5016280
     B      0.5000038  0.5000038  0.0016201
     B      0.5000038  0.5000038  0.5016280
     B      0.2500019  0.2500019  0.2516239
     B      0.2500019  0.2500019  0.7516319
     B      0.2500019  0.7500057  0.2516239
     B      0.2500019  0.7500057  0.7516319
     B      0.7500057  0.2500019  0.2516239
     B      0.7500057  0.2500019  0.7516319
     B      0.7500057  0.7500057  0.2516239
     B      0.7500057  0.7500057  0.7516319
     N      0.0000000  0.0000000  0.3434755
     N      0.0000000  0.0000000  0.8434833
     N      0.0000000  0.5000038  0.3434755
     N      0.0000000  0.5000038  0.8434833
     N      0.5000038  0.0000000  0.3434755
     N      0.5000038  0.0000000  0.8434833
     N      0.5000038  0.5000038  0.3434755
     N      0.5000038  0.5000038  0.8434833
     N      0.2500019  0.2500019  0.0934715
     N      0.2500019  0.2500019  0.5934794
     N      0.2500019  0.7500057  0.0934715
     N      0.2500019  0.7500057  0.5934794
     N      0.7500057  0.2500019  0.0934715
     N      0.7500057  0.2500019  0.5934794
     N      0.7500057  0.7500057  0.0934715
     N      0.7500057  0.7500057  0.5934794
     H      0.0000000  0.4259533  0.3095699
     H      0.0000000  0.4259533  0.8095778
     H      0.0000000  0.9259571  0.3095699
     H      0.0000000  0.9259571  0.8095778
     H      0.5000038  0.4259533  0.3095699
     H      0.5000038  0.4259533  0.8095778
     H      0.5000038  0.9259571  0.3095699
     H      0.5000038  0.9259571  0.8095778
     H      0.2500019  0.3240524  0.0595660
     H      0.2500019  0.3240524  0.5595738
     H      0.2500019  0.8240562  0.0595660
     H      0.2500019  0.8240562  0.5595738
     H      0.7500057  0.3240524  0.0595660
     H      0.7500057  0.3240524  0.5595738
     H      0.7500057  0.8240562  0.0595660
     H      0.7500057  0.8240562  0.5595738
     H      0.0000000  0.0999007  0.0389506
     H      0.0000000  0.0999007  0.5389586
     H      0.0000000  0.5999046  0.0389506
     H      0.0000000  0.5999046  0.5389586
     H      0.5000038  0.0999007  0.0389506
     H      0.5000038  0.0999007  0.5389586
     H      0.5000038  0.5999046  0.0389506
     H      0.5000038  0.5999046  0.5389586
     H      0.2500019  0.1501012  0.2889545
     H      0.2500019  0.1501012  0.7889625
     H      0.2500019  0.6501050  0.2889545
     H      0.2500019  0.6501050  0.7889625
     H      0.7500057  0.1501012  0.2889545
     H      0.7500057  0.1501012  0.7889625
     H      0.7500057  0.6501050  0.2889545
     H      0.7500057  0.6501050  0.7889625
     H      0.4316033  0.0395003  0.3095049
     H      0.4316033  0.0395003  0.8095127
     H      0.4316033  0.5395041  0.3095049
     H      0.4316033  0.5395041  0.8095127
     H      0.9316071  0.0395003  0.3095049
     H      0.9316071  0.0395003  0.8095127
     H      0.9316071  0.5395041  0.3095049
     H      0.9316071  0.5395041  0.8095127
     H      0.0684005  0.0395003  0.3095049
     H      0.0684005  0.0395003  0.8095127
     H      0.0684005  0.5395041  0.3095049
     H      0.0684005  0.5395041  0.8095127
     H      0.5684043  0.0395003  0.3095049
     H      0.5684043  0.0395003  0.8095127
     H      0.5684043  0.5395041  0.3095049
     H      0.5684043  0.5395041  0.8095127
     H      0.3184024  0.2105016  0.0595009
     H      0.3184024  0.2105016  0.5595089
     H      0.3184024  0.7105054  0.0595009
     H      0.3184024  0.7105054  0.5595089
     H      0.8184062  0.2105016  0.0595009
     H      0.8184062  0.2105016  0.5595089
     H      0.8184062  0.7105054  0.0595009
     H      0.8184062  0.7105054  0.5595089
     H      0.1816014  0.2105016  0.0595009
     H      0.1816014  0.2105016  0.5595089
     H      0.1816014  0.7105054  0.0595009
     H      0.1816014  0.7105054  0.5595089
     H      0.6816052  0.2105016  0.0595009
     H      0.6816052  0.2105016  0.5595089
     H      0.6816052  0.7105054  0.0595009
     H      0.6816052  0.7105054  0.5595089
     H      0.0866007  0.4505034  0.0385006
     H      0.0866007  0.4505034  0.5385085
     H      0.0866007  0.9505072  0.0385006
     H      0.0866007  0.9505072  0.5385085
     H      0.5866045  0.4505034  0.0385006
     H      0.5866045  0.4505034  0.5385085
     H      0.5866045  0.9505072  0.0385006
     H      0.5866045  0.9505072  0.5385085
     H      0.4134031  0.4505034  0.0385006
     H      0.4134031  0.4505034  0.5385085
     H      0.4134031  0.9505072  0.0385006
     H      0.4134031  0.9505072  0.5385085
     H      0.9134070  0.4505034  0.0385006
     H      0.9134070  0.4505034  0.5385085
     H      0.9134070  0.9505072  0.0385006
     H      0.9134070  0.9505072  0.5385085
     H      0.1634012  0.2995023  0.2885046
     H      0.1634012  0.2995023  0.7885125
     H      0.1634012  0.7995061  0.2885046
     H      0.1634012  0.7995061  0.7885125
     H      0.6634051  0.2995023  0.2885046
     H      0.6634051  0.2995023  0.7885125
     H      0.6634051  0.7995061  0.2885046
     H      0.6634051  0.7995061  0.7885125
     H      0.3366026  0.2995023  0.2885046
     H      0.3366026  0.2995023  0.7885125
     H      0.3366026  0.7995061  0.2885046
     H      0.3366026  0.7995061  0.7885125
     H      0.8366064  0.2995023  0.2885046
     H      0.8366064  0.2995023  0.7885125
     H      0.8366064  0.7995061  0.2885046
     H      0.8366064  0.7995061  0.7885125
\end{verbatim}
\end{center}

\newpage\section{Detailed Rotational Barriers}

\mbox{}

\begin{figure}[h]
\centering
\includegraphics[width=\columnwidth]{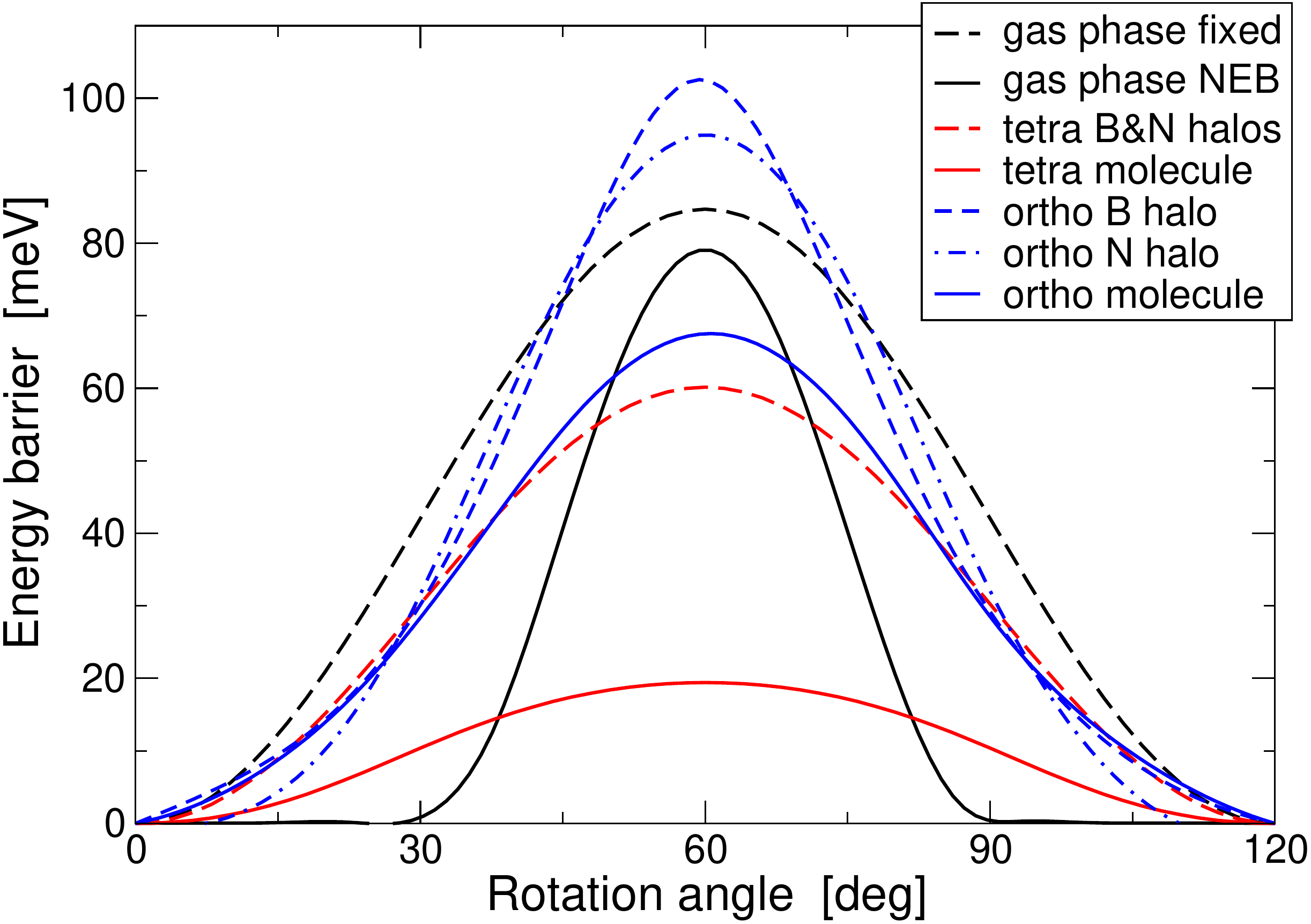}
\caption{\label{fig:barriers}Comparison of rotational barriers as
obtained by NEB calculations and fixed-geometry calculations. The maxima
of these curves are tabulated in Table~1 of the main manuscript.}
\end{figure}

\vspace*{3.5in}

\bibliography{references}
